\documentclass[twocolumn,nofootinbib,preprintnumbers,amsmath,amssymb]{revtex4}

\usepackage{graphics,graphicx,epsfig}
\usepackage{amsfonts}
\usepackage{ifthen}
\usepackage[dvips]{color}
\usepackage{latexsym}
\usepackage{slashed}
\usepackage{mathrsfs}

\newcounter{mysubequation}[equation]

\newcommand{\be}{\begin{equation}}
\newcommand{\ee}{\end{equation}}
\newcommand{\bea}{\begin{eqnarray}}
\newcommand{\eea}{\end{eqnarray}}
\newcommand{\bean}{\begin{eqnarray*}}
\newcommand{\eean}{\end{eqnarray*}}

\newcommand{\G}{\Gamma}
\newcommand{\T}{\Theta}

\newcommand{\N}{{\cal N}}
\newcommand{\V}{{\cal V}}

\newcommand{\IP}{{\mathbb P}}

\newcommand{\IZ}{{\mathbb Z}}

\def\eg{{\eg e.g.}}

\newcommand{\lsim}{\,\raise.3ex\hbox{$<$\kern-.75em\lower1ex\hbox{$\sim$}}\,}
\newcommand{\gsim}{\,\raise.3ex\hbox{$>$\kern-.75em\lower1ex\hbox{$\sim$}}\,}

\def\Ka{K\"{a}hler}

\def\nK{non-K\"{a}hler}

\begin{document}
\preprint{hep-th/0703048}
\preprint{MIT-CTP 3816}
\title{Conformal Field Theory and the Reid Conjecture}

\author{Allan Adams}
\affiliation{Center for Theoretical Physics, Massachusetts Institute of Technology, Cambridge, MA  02139 USA}

\date{\today}

\begin{abstract}
\noindent
We construct special pairs of quantum sigma models on K\" ahler Calabi-Yau and non-K\" ahler Fu-Yau manifolds which flow to the same conformal field theories in their ``small-radius'' phases.
This smooth description of a novel type of topology change constitutes strong evidence for Reid's conjecture on the connectedness of moduli spaces of \Ka\ and \nK\ manifolds with trivial canonical class.
\end{abstract}

\maketitle

\section{Introduction}
\setcounter{equation}{0}

Do all Calabi-Yau manifolds live on the same moduli space?  In two complex dimensions, the answer is yes: all K3 manifolds, whether built out of hypersurfaces or orbifolds or glued-together macaroni, may be smoothly deformed into each other by the continuous variation of metric moduli.  The fact that all K3s
share the same Hodge numbers makes this classically plausible.

The case of Calabi-Yau 3-folds remains far more mysterious.  Since a generic pair of Calabi-Yaus do not share the same Hodge numbers, any putatively ``smooth'' transition between them must involve rather dramatic kind of ``smooth'' topology change.  In a very beautiful paper \cite{Strominger:Conifold}, Strominger explained why this is not as much of an obstruction as one might think: two topologically inequivalent Calabi-Yaus may be connected by a local conifold transition which, while of course geometrically singular as it involves a local surgery \cite{Surgery}, may be completely smooth in non-perturbative string theory due to the condensation of extremal black holes.  These massless black holes were subsequently reinterpreted as branes wrapping the shrinking cycle in \cite{GMS}.  By working in an appropriate quantum version of the geometry, then, such transitions may well be smooth in some appropriate sense.

It is thus natural to wonder whether the moduli space of Calabi-Yaus is in fact smoothly connected via local conifold transitions.  Of course, since a local conifold preserves $h^{1,1}-h^{2,1}$, not all CYs may be directly connected by a simple local conifold.  More significantly, a simple argument (see eg \cite{Reid} and \cite{WuYen} and references therein) demonstrates that going through a local conifold transition on a  generic compact Calabi-Yau does {\em not} result in another Calabi-Yau:  the resulting manifold again has trivial canonical class but is {\em not}, in general, \Ka.  

Such arguments lead to a natural conjecture, and a rather more spectacular fantasy, generally referred to as the Reid conjecture \cite{Reid}.  The minimal conjecture is that the moduli space of Calabi-Yaus is connected to the moduli space of \nK\ manifolds with trivial canonical class via local conifold transitions which, in the appropriate category\footnote{Which, as we shall argue, may be conformal field theory.}, should be smooth.  The fantasy is that all Calabi-Yaus are connected via such transitions.

In this paper we shall construct smooth transitions between special families of \Ka\ and \nK\ manifolds with trivial canonical class by showing that the (0,2) superconformal field theories describing both geometries become isomorphic in certain small-radius phases.  Moreover, the transition between the two descriptions will involve a very special kind of surgery, essentially bosonization and fermionization, which is explicitly non-local in the spacetime description but completely smooth in the CFT.  Interestingly, a straightforward 
argument suggests that these transitions should not be possible in the worldsheet theory of the Heterotic string on these compactifications, but only in a ``probe'' CFT uncoupled to (0,1) worldsheet supergravity, as we shall see below.

To get a controlled description of these worldsheet theories it is particularly convenient to use the gauged linear sigma model description of \nK\ torsion\footnote{A compact complex \nK\ 3-fold with special holonomy necessarily has intrinsic torsion.} compactifications developed in \cite{Adams:2006kb}, to which we refer the reader for a detailed introduction.  The structure of these models differs from that of more familiar gauged linear sigma models in two important respects.  First, the fermions transform in chiral representations of the gauge group, leading to non-vanishing 1-loop gauge anomalies proportional to $ch_{2}(T_{S})-ch_{2}(\V_{S})$, where $S=K3$, the ``Fu-Yau'' compactification\footnote{The basic structure of this compactification has appeared in numerous forms with different levels of elaboration over the years.  The ansatz first appeared via duality in \cite{Sav} and was elaborated in \cite{BD}. A tour de force proof that this ansatz solves the Strominger equations \cite{Strominger:Torsion}\ first appeared in \cite{FuYau}, with further explication in \cite{BBFTY}. For a detailed history of the early literature, see 
 \cite{Mike}.}, will be our basic example.  Secondly, these models contain novel (0,2) multiplets whose lowest components are doublets of gauge-charged axions coordinatizing the $T^{2}$ fibres over $S$, with the classical gauge-variation of the axion action precisely canceling the 1-loop anomaly in a 2d avatar of the Green-Schwarz mechanism.  If we further fix to rectangular complex structure on the $T^{2}$ fibres, (0,2) supersymmetry constrains the radii of the two circles to lie at (flux-quantized multiples of) the self-dual radius.

We begin by constructing a pair of worldsheet theories of the above form and identifying a non-trivial relation between the two, then explain why this worked and how to generalize it to a large class of other examples.

\section{A Mischievous Duality}

Consider a (2,2) GLSM for the sextic $K3$ hypersurface in $W\IP_{1122}$, expressed in (0,2) formalism.  The model includes\footnote{The model also contains a single uncharged scalar $\Sigma$ which contributes right-moving fermions playing the role of the right-moving gauginos required by (2,2) supersymmetry, though it will play no role in what follows.} 
a single $U(1)$ vector coupled to five chirals $\{P, \Phi_{i}\}$ and five Fermi multiplets $\G_{i}$ with charge assignments
$$
\begin{tabular}{ccccc|ccccc}
$P$   &$\Phi_{1}$&$\Phi_{2}$&$\Phi_{3}$ &$\Phi_{4} $&$\G_{P}$ &$\G_{1}$ &$\G_{2}$ &$\G_{3}$ &$\G_{4}$  \\
-6  &1             &1             &2             &2             &-6         &1          &1          &2          &2         
\end{tabular}
$$
To get a complex 3-fold we can simply tack on a free $T^{2}$, which we can incorporate by adding a free (0,2) torsion multiplet $\T$ (whose two real scalars are compact) and a free fermi multiplet $\G_{\T}$ to fill out a (2,2) multiplet.  The Higgs branch of the theory is thus a simple product, $X=T^{2}\times K3$, where our $K3$ is singular, since any transverse sextic necessarily intersects the $\IZ_{2}$-orbifold fixed-curve of the ambient $W\IP_{1122}$, $\phi_{1}=\phi_{2}=0$.

This $\IZ_{2}$ singularity begs to be resolved, so let's blow it up in a (2,2)-symmetric fashion, replacing the point $\phi_{1}=\phi_{2}=0$ with a $\IP^{1}$.   To do this, we add an additional $U(1)$ whose D-term will enforce $|\phi_{1}|^{2}+|\phi_{2}|^{2} = r > 0$, and add a (chiral,fermi) pair $(\wp,\G_{\wp})$ to provide a coordinate on the exceptional divisor.  Including the decoupled $T^{2}$ multiplets for fun, the resulting model has charge assignments
$$
\begin{tabular}{ccccc|ccccc||cc||cc}
$P$   &$\Phi_{1}$&$\Phi_{2}$&$\Phi_{3}$ &$\Phi_{4} $&$\G_{P}$ &$\G_{1}$ &$\G_{2}$ &$\G_{3}$ &$\G_{4}$ &$\wp$ &$\G_{\wp}$ &$\Theta$ &$\G_{\Theta}$ \\
-6  &1             &1             &2             &2             &-6         &1          &1          &2          &2          &0          &0                   &0     &0\\
0  &1             &1             &0             &0             &0         &1          &1          &0          &0          &-2          &-2                   &0     &0
\end{tabular}.
$$
Since each right-moving fermion comes paired with a left-moving fermion with the same charge, the chiral anomaly vanishes, as it must by (2,2) susy.  The total space is thus $\tilde{X}=T^{2}\times\tilde{K3}$, a free $T^{2}$ times the smooth sextic $K3$ in resolved $\tilde{W\IP}_{1122}$.  

This model has a closely related cousin with identical field content but slightly different charge assignments,
$$
\begin{tabular}{ccccc|ccccc||cc||cc}
$P$   &$\Phi_{1}$&$\Phi_{2}$&$\Phi_{3}$ &$\Phi_{4} $&$\G_{P}$ &$\G_{1}$ &$\G_{2}$ &$\G_{3}$ &$\G_{4}$ &$\wp$ &$\G_{\wp}$ &$\Theta$ &$\G_{\Theta}$ \\
-6  &1             &1             &2             &2             &-6         &1          &1          &2          &2          &0          &0                   &0     &0\\
0  &1             &1             &0             &0             &0         &1          &1          &0          &0          &-2          &0                   &-2(1+i)     &0
\end{tabular}.
$$
The fermi multiplet $\G_{\wp}$ is now gauge-neutral, so the fermion spectrum is chiral and the second $U(1)$ is anomalous.   Meanwhile, the torsion multiplet is now gauge-charged, so the classical action is no longer gauge invariant; instead, this classical variation precisely cancels the one-loop anomaly by a 2d Green-Schwarz mechanism.  For large values of the two FI parameters, this model flows to a non-linear sigma model on a \nK\ $T^{2}$-fibration over the (resolved) $\tilde{K3}$ with non-trivial torsion supported entirely over the exceptional divisor.  This is a complex 3-fold of the type studied in \cite{FuYau}.

Here's the remarkable thing.  Take the FI parameter of the second $U(1)$ large and negative, $r^{2}\ll -1$.  This forces $\wp\neq0$, breaking the second $U(1)$ to a $\IZ_{2}$ subgroup.  In both gauge theories, this $\IZ_{2}$ only acts non-trivially\footnote{Of course, there is a $\IZ_{2}$ subgroup of the first $U(1)$ which acts identically, so, in both theories, there is a linear combination of the two which is completely ineffective.  It's straightforward to deal with this by rediagonalizing the generators of the gauge group, but, since it will make no difference for anything we do, we'll just ignore this detail and press on.} on $(\Phi_{1,2},\G_{1,2})$; in particular, it leaves $(\wp,\G_{\wp})$ and $(\T,\G_{\T})$ invariant.  In the phase $r^{2}\ll-1$, then, both models flow to the same conformal field theory!

Of course, the mixed phase $r^{1}\gg1$, $r^{2}\ll-1$ is a slightly odd beast and should be treated with some care, since we do not have a good weakly coupled description of the full theory beyond the chiral sector.  To nail down the connection between these two models in a completely controlled regime, it is helpful to drive {\em both} FI parameters to large negative values, which takes both models to simple Landau-Ginzburg orbifold CFTs.  It follows from the above that the two LG orbifolds are isomorphic.

This is actually a very common feature of (0,2) model building first pointed out in the very beautiful work of \cite{ShamitLG}, where it was argued that two Heterotic compactifications on the inequivalent Calabi-Yaus with inequivalent gauge bundles may nonetheless flow to the same Landau-Ginzburg model deep in the interior of their \Ka\ moduli spaces.  In many (though by no means all) cases, it is known that multiple large-radius phases are connected to the same LG model by marginal flows in inequivalent gauged linear sigma models.  The construction above represents a simple generalization of this construction to the context of \nK\ compactifications with intrinsic torsion.

In this torsion case, interestingly, we can say slightly more.  Consider the \Ka\ model describing $K3\times T^{2}$ and take both legs of the $T^{2}$ to the free-fermion radius, with rectangular complex structure.  This lets us do something very entertaining: we can fermionize 
%
%
%
%
%
%
%
%
the torsion multiplet to give a free complex fermion and simultaneously bosonize the complex fermion\footnote{This is usually done with gauge-neutral fields; however, requiring bosonization to reproduce both the UV free field OPEs and the chiral gauge anomaly uniquely specifies the bosonized theory.  That the gauging of the scalars in the torsion multiplet is chiral (see \cite{Adams:2006kb} for a discussion of the torsion multiplet) dovetails naturally with the anomalous gauging of the fermions.} $\G_{\wp}$ to give a pair of $S^{1}$-valued scalars at free-fermion radius coupled to the gauge field with charge -2.  As it happens, free-fermion radius is exactly where a torsion multiplet with charge $-2(1+i)$ is required to live by (0,2) supersymmetry.  But this is nothing other than the \nK\ model studied above at a special point\footnote{Though not the Landau-Ginzburg point -- it is to a special point in the complex structure moduli space of the $T^{2}$ that we've fixed.  We could of course change coordinates so that only one circle is charged, placing that circle at the self-dual radius; we secretly chose the charge assignments precisely to fix these radii.}  in its moduli space!

This change of variables is less mysterious in the phase $r^{2}\ll -1$, where the exceptional divisor is blown down and the second $U(1)$ is broken to its $\IZ_{2}$ subgroup, leaving both $\G_{\wp}$ and $\T$ free.  Bosonization and fermionization are then completely trivial, ie the ``surgery'' connecting the \Ka\ and \nK\ geometries is naturally performed when the relevant cycles are blown down.  That said, even in this free limit, the requisite bosonization and fermionization mix (0,1) multiplets with reckless disregard for 
the {\em particular} (0,1) we would have gauged had we coupled our matter theory to worldsheet supergravity to build a Heterotic string\footnote{Particular thanks to J.~McGreevy for emphasizing this point.}.  This suggests that the transition above, while realizable in the conformal field theory to which the linear models flow, is not realizable by tuning mutually local marginal vertex operators in the worldsheet theory of the Heterotic string on these compactifications.

It is completely straightforward to use the torsion gauged linear sigma model to construct many more families of \Ka\ and \nK\ compactifications along these lines.  The basic strategy boils down to ensuring that all non-trivial torsion and non-standard-embedding of the gauge bundle is supported entirely on some exceptional $\IP^{1}$s; by constructing pairs of models with either non-trivial torsion, or non-trivial vector bundle, over only the exceptional divisor, the resulting worldsheet theories will  become equivalent, up to possibly anomalous discrete orbifoldings, when the exceptional divisor is blown down, or, analogously, at the Landau-Ginzburg point of the base $K3$.  The study of the resulting asymmetric Landau-Ginzburg orbifolds is extremely interesting in its own right, and will be reported on elsewhere \cite{LGPaper}; the example above was constructed specifically so as to avoid such technical complications.

\section{Connectedness and Other Fantasies}

In the above we studied special pairs of \Ka\ and \nK\ manifolds with special holonomy and argued that both have classically singular orbifold limits which are controlled by the same (0,2) supersymmetric conformal field theory.  At the level of conformal field theory, then, the moduli space of at least some Calabi-Yau manifolds is smoothly connected to the moduli space of certain topologically inequivalent \nK\ manifolds with trivial canonical class.  Moreover, the transition involves blowing down a $\IP^{1}$ in the Calabi-Yau, then performing a special kind of surgery -- here, fermionization and bosonization in the CFT -- to land on a \nK\ manifold in which an $S^{1}$ is fibred over the (erstwhile) $\IP^{1}$ via the Hopf fibration to give, topologically, an $S^{3}$.  This surgery becomes a trivial matter when the $\IP^{1}$ ($S^{3}$) is blown down; this is where the conformal field theories become trivially isomorphic.   Taking both descriptions to Landau-Ginzburg points similarly gives the same Landau-Ginzburg models, providing a particularly well-behaved, computationally tractable description of the conformal field theory of this torsion compactifications.

Many of the arguments presented above depend crucially on the well-posedness of the gauged linear sigma model description of the torsion compactifications.  While we believe this to be a very reliable and well-behaved model, there are certainly many open questions in the study of these models and the conformal theories to which they flow, so it is worth pointing out that many features of these models do {\em not} depend on the detailed structure of the linear model.  Most importantly, the construction of the torsion compactification over the sextic in $W\IP_{1122}$ with torsion supported entirely over the exceptional divisor can be studied directly at the level of the 1-loop spacetime supergravity; once the exceptional divisor is blown down, the geometry is indistinguishable from the blown-down \Ka\ geometry.  This should persist in any description for which the 1-loop supergravity analysis is a reliable approximation.

Note that the worldsheet gauge anomaly was crucial in our construction.  In Type II such a transition simply cannot occur without the introduction of additional non-perturbative degrees of freedom -- cf Strominger's beautiful description of the condensation of wrapped branes on the conifold.  Only with chiral $\N$=(0,2) supersymmetry do we have both the flexibility and the control needed to follow such a transition purely from the worldsheet perspective.  Additionally, the transition involves a non-local change of spacetime coordinates mixing bundle and metric degrees of freedom -- something known to happen under mirror symmetry in the Heterotic string \cite{MS:02}, but nonetheless extremely mysterious.   Relatedly, this is deeply at odds with the usual coupling of the matter-sector CFT to worldsheet gravity, as we have seen.

Note too\footnote{Special thanks to Shamit Kachru for discussions on these points.} that this construction smoothly relates inequivalent large radius geometries while avoiding certain singularities which sometimes appeared in the very beautiful dualities of \cite{ShamitLG}.  To ensure this, it was extremely helpful to start with a model inheriting the singularity structure of a (2,2) model and add a smooth (in particular, free) structure over the blown-down exceptional divisor.  The sextic example of Section 2 was also particularly simple in its orbifold and Landau-Ginzburg orbifold phases.  In more general examples, the orbifold action acts non-trivially on the exceptional fermi and torsion multiplets, leading to an anomalous LG orbifold whose anomaly is cancelled by inflow from a gauged WZW model describing the $T^{2}$ torsion multiplet.  (These more general models will be described in a forthcoming note \cite{LGPaper}.)  For the special case discussed above, the anomaly is trivial at the LG point, which incidentally provides an independent argument for the Gepner model arising in the moduli space of the sextic in $W\IP_{1122}$ which doesn't rely on the normal relations between LG monomials and minimal models.

One obvious lacuna in our discussion is a description of a transition in a general Calabi-Yau 3-fold which is not a direct product $K3\times T^{2}$, ie with irreducible holonomy.  While our construction does not directly generalize, there is a natural strategy to extend our discussion to elliptically fibred Calabi-Yau near a point in its moduli space where some set of $\IP^{1}$s in the base degenerate.  Near each degeneration, the Calabi-Yau is well described by a local model of the form studied in this paper (though with more general base $S$); performing our CFT surgery at each degeneration then gives a set of local torsion models which should patch together to give a compact global torsion compactification.  To connect  to the mathematical model of these transitions \cite{Surgery,Reid}, it remains to be shown that these local torsion models can indeed be patched together.  

A complementary approach to studying Heterotic transitions between \Ka\ and \nK\ geometries involves studying the spacetime supergravity, where such a transition would appear as a geometric transition involving Heterotic fivebranes wrapping shrinking cycles in the \Ka\ geometry.  While this is morally similar to the type II transition, it is considerably more demanding technically, largely due to the complexity of the Heterotic fivebrane worldvolume theory, but also because of the difficulty of studying the Heterotic theory on connected sums of $S^{3}\times S^{3}$s.  This approach is being developed in detail by the authors of \cite{FuYau,BBFTY} (see \cite{BBFTY:WIP}), and should shortly provide a powerful set of tools for studying the geometry directly.  Connecting the resulting spacetime description to this conformal field theory analysis should prove extremely interesting.

This construction leaves several obvious questions largely or completely unanswered.  For example, it would be very interesting to know whether generic topologically inequivalent Calabi-Yaus may be smoothly connected via excursions onto the moduli space of \nK\ manifolds.   At first blush, from the point of view of the linear model, this looks extremely likely, since all that is required is a \nK\ manifold with more than one singular degeneration where the torsion multiplet becomes free; we are currently hunting such a snark.  The obvious next question is whether the universal moduli space of manifolds of special holonomy, of which the moduli space of Calabi-Yaus is a disconnected submanifold, is connected.   As yet we have nothing sharp to say one way or the other.  However, the results of this note strongly suggest that the way forward is Heterotic.

\vfill
\begin{acknowledgments}
\noindent 
I would like to thank 
M.~Becker,
S.~Kachru,
A.~Lawrence,
D.~Morrison,
S.~Sethi,
D.~Tong,
L.-S.~Tseng,
S.-T.~Yau
and especially
J.~Lapan and
J.~\mbox{McGreevy}
for many fun and enlightening conversations, with particular thanks to J.~McG. for encouraging me to write up these observations.  Thanks also to the organizers and participants of the Amsterdam Summer String Workshop and KITP {String Phenomenology} Program, where some of this work was discussed.  This work was supported in part by the DOE under contract No.~DE-FC02-94ER40818.
\end{acknowledgments}

\end{document}